\documentclass[english,12pt,preprint]{aastex}

\usepackage{babel}

\shorttitle{\indent \def Multi-strand coronal loop}
\shortauthors{Bourouaine and Marsch}

\begin{document}

\title{Multi-strand coronal loop model and filter-ratio analysis}

\author{Sofiane Bourouaine and Eckart Marsch}

\affil{Max-Planck-Institut f\"{u}r Sonnensystemforschung, 37191
Katlenburg-Lindau, Germany}

\begin{abstract}
We model a coronal loop as a bundle of seven separate strands or filaments.
Each of the loop strands used in this model can independently be heated (near
their left footpoints) by Alfv\'en/ion-cyclotron waves via wave-particle
interactions. The Alfv\'en waves are assumed to penetrate the strands from
their footpoints, at which we consider different wave energy inputs. As a
result, the loop strands can have different heating profiles, and the
differential heating can lead to a varying cross-field temperature in the
total coronal loop. The simulation of TRACE observations by means of
this loop model implies two uniform temperatures along the loop length,
one inferred from the 171:195 filter ratio and the other from the
171:284 ratio. The reproduced flat temperature profiles are consistent
with those inferred from the observed EUV coronal loops. According to our
model, the flat temperature profile is a consequence of the coronal loop
consisting of filaments, which have different temperatures but almost
similar emission measures in the cross-field direction. Furthermore,
when we assume certain errors in the simulated loop emissions (e.g., due to
photometric uncertainties in the TRACE filters) and use the triple-filter
analysis, our simulated loop conditions become consistent with those of
an isothermal plasma. This implies that the use of TRACE/EIT triple filters
for observation of a warm coronal loop may not help in determining
whether the cross-field isothermal assumption is satisfied or not.
\end{abstract}

\keywords{Sun: solar corona, waves, Sun: magnetic fields, particle acceleration}

\email{bourouaine@mps.mpg.de}

\section{Introduction}

In our recent papers \citet{Bourouaine2008a,Bourouaine2008b} (hereafter
referred to as BR2008a; BR2008b) we modelled a coronal loop as one single
flux tube (or monolithic loop) in which the confined plasma is heated by
ion-cyclotron/Alfv\'en waves via the resonant wave absorption mechanism. When
considering Coulomb collisions in that model, it turned out that they are not
strong enough to maintain local thermal equilibrium, and it was found that
the electron temperature profile is different from the ion profile.
Furthermore, it was shown that the warm coronal loop (e.g., $T\leq 1.5$ MK),
which has a quasi-uniform electron temperature profile and enhanced density
\citep[relative to the static model by][]{Serio1981}, is the consequence of a
uniform cross-section of the flux tube forming the loop. Modelling of this
type of loop profile with a uniform temperature was motivated by the many
observations of EUV (Extreme Ultraviolet) loop emissions (using the high
spatial resolution imaging provided by TRACE
\citep{Lenz1999,Aschwanden1999,Aschwanden2000}.

Under the assumption that the loops have a single cross-field temperature,
techniques based on filter-ratio analysis have been employed to infer the
electron temperature along the EUV coronal loops
\citep{Noglik2007,Noglik2008,Aschwanden2008,Schmelz2009}. Most of the
obtained results indicate that the warm loops have roughly constant
temperatures along much of their lengths. Furthermore, it was found that
their densities are roughly uniform and much larger than those predicted by
static models with uniform heating \citep{Aschwanden2001,Winebarger2003}.
However, it was concluded that static loop models with footpoint heating
may produce flat temperature profiles and enhanced apex densities,
but in the case of long loops the theoretical enhancement in density
was found to be not large enough as compared to the observed values.

It is believed that the difficulty of reproducing the observed properties of
TRACE loops with static models lies in that they are not in thermal
equilibrium. Therefore, it was suggested that impulsive heating caused by
nanoflares could explain the observed loop characteristics. However, in such
process, numerical simulations showed that the single-loop cooling would
appear simultaneously in both the TRACE 171~{\AA} and 195~{\AA} filters for a
few seconds \citep{Reeves2002}. This cooling time is much shorter than the
lifetime of the active-region loops observed by TRACE, which generally live
for several hours while maintaining their high densities and flat temperature
profiles.

\begin{figure}[t]
\includegraphics[width=14cm,height=5.5cm]{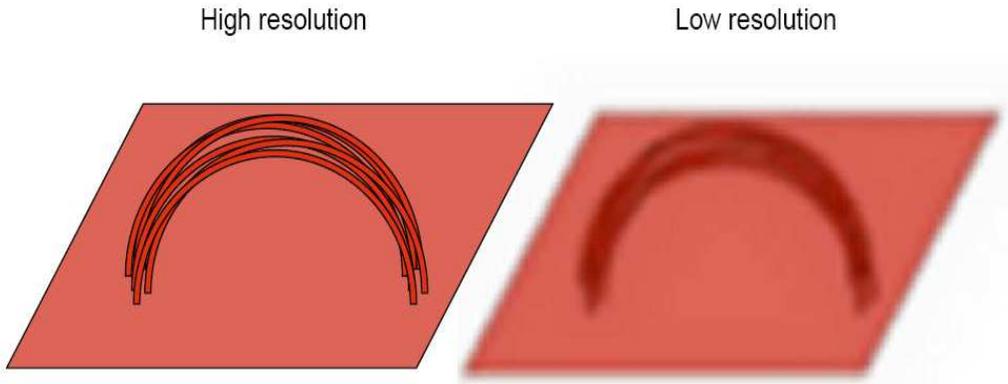}
\caption{Coronal model loop that consists of a bundle of many magnetic strands}
\label{fig.1}
\end{figure}

Therefore, it has been suggested that the relevance of the impulsive heating
heavily relies on the assumption of a multi-strand or multi-thread nature of
coronal loops, as it is illustrated in Fig.~\ref{fig.1}, and these strands
are at different stages of heating and cooling exhibiting a broad
differential emission measure (DEM) distribution.

The concept of multi-stranded loops is supported
by many observations which exploit the high spatial resolution of TRACE and
indicate that an observed "fat" single loop could be composed of many
small-scale filaments having most probably different temperatures across the
magnetic field \citep{Schmelz2001,Schmelz2003,Schmelz2005,Martens2002}.
Hydrodynamic models based on multi-stranded loops have been suggested and
used to calculate numerically the flare emission \citep{Warren2006, Reeves2007}.
Other multi-strand coronal loop models involving nanoflare heating have been
proposed to simulate EUV coronal loops
\citep[e.g.,][]{Cargill1997, Warren2002, Sarkar2009}.

The lack of observations of very fine-scale structures at scales below
the TRACE resolution still permits the possibility that the observed loops
may be composed of very thin "strands or filaments" which have different
temperature and density profiles. As a result, the observed loop intensity
could be caused by an ensemble of emissions from many unresolved strands.

\citet{Reale2000} produced the observed features of warm coronal loops by
assuming them to consist of a bundle of static, uniformly heated strands.
Their model, however, implements dense and hot strand ($\sim5$~MK) to
reproduce the flat TRACE filter ratios. This seems to contradict previous
observations which indicate that, generally, hot loops are not co-spatial
with relatively cool loops \citep{Sheeley1980,Habbal1985, Brooks2007}.

Following the concept of a multi-strand structure of coronal loops, we
will also model the observed TRACE loop as a bundle of seven isolated
strands. However, concerning loop heating we propose here that the
strands are heated through the dissipation of parallel propagating
ion-cyclotron/Alfv\'{e}n waves via resonant wave absorption. The same
kinetic model that has been used in the papers BR2008a and BR2008b
to describe the heating of a monolithic loop is now applied to the
heating of an individual small-scale strand. Thus various fine structures
can have different temperatures and density profiles. Therefore, an
observation of a neighbouring set of filaments with a low-resolution
instrument could lead to an apparently fat loop, which yet is simply
composed of a finite number of strands. Since in the corona the cross-field
heat transport is very weak, and as the plasma is dominated by the magnetic
pressure, we can assume that separate loop strands are thermally isolated.

Once we have determined the different plasma profiles of single loop strands,
we can synthesize the total emission of the composite model loop, thereby
assuming its observation as a monolithic loop in the TRACE/EIT filters, and
thus we can extract its temperature by adopting the filter-ratio technique.
As a result we find that, if the coronal loop is composed of unresolved filaments
having different temperatures and roughly identical emission measures across
the coronal loop, a quasi-uniform temperature is inferred along the loop length.

The paper is organized as follows: in Sec.~2 we model a TRACE/EIT coronal
loop as a bundle of seven separate filaments heated through the dissipation
of high-frequency Alfv\'en waves. We assume different wave energy inputs at
the footpoints of the different strand. Then in Sec.~3, we synthesize the
emission of the modelled coronal loop in the three TRACE passbands, and thus
we can derive the loop temperature from two filter ratios 171:195 and 171:284.
Finally, we discuss the obtained results and conclude in Sec.~4.

\section{Multi-stranded coronal loop model}

It is widely believed that the small-scale reconnection events occurring
in the lower corona (or chromosphere) can be a source of high-frequency
plasma waves, namely waves having frequencies comparable to the local
ion-gyrofrequency \citep{Axford1992,Axford1999}. But only recently,
by using STEREO, Helios and Cluster data, have ion-cyclotron waves clearly
been detected at different distances in the solar wind \citep[see][]{Jian2008}.
Previously, theoretical fluid models \citep{Marsch1997,Tu1997,Hackenberg2000,He2008}
have shown that the propagation of such waves into the corona and
their subsequent absorption can lead to efficient ion heating in open magnetic
structures (such as coronal funnels). Thus kinetic models could explain the rapid
temperature increase in the transition region and corona \citep{Vocks2002a,Vocks2002b}.
Therefore, this wave heating mechanism may also be relevant to the
closed structures like coronal loops. Indeed, that formation of hot
coronal loops can be a consequence of the dissipation of ion-cyclotron
waves along the closed magnetic fields confining the plasma in the
loops, was shown by \citet{Bourouaine2008a, Bourouaine2008b} using
kinetic models.

Furthermore, intuitively we may think that the random spatial occurrence of
reconnection events may generate ion-cyclotron waves with different amounts
of energy at various locations. The dissipation of the waves along
neighbouring closed field lines originating there can lead to plasma heating
and acceleration on those field lines, thus producing small-scale plasma
filaments or magnetic loop strands (which may in the extreme case have scales
comparable to the ion-skin depth but certainly below the the present TRACE
spatial resolution). These filaments may have differential temperature and
density profiles, and the observations of neighbouring sets of such filaments
with low-resolution instruments may reveal an apparent fat coronal loop that
really is composed of a finite number of those small-scale strands (see
Fig.~\ref{fig.1}).

In this paper, we will represent a coronal loop as an ensemble of
seven separated loop strands or filaments. Each strand is modelled
as an electron-proton plasma confined within a semi-circular cylindric
and symmetric flux tube. All the strands (numbered by index $i$) are
assumed to have the same length ($L=63$~Mm) and the same local width,
$w_{i}(s)$. We suggest that these strands are close to each other and situated
in planes that are perpendicular to the solar surface. The flux tubes of these
strands expand similarly with height, i.e., the filaments have the
same expansion factor, $\Gamma=1.48$. This means that each strand has
a varying cross-section which expands from the strand footpoints (situated
at the transition region) to the strand top and satisfies formula
(13) in the paper BR2008a.

The loop strands are heated by dissipation of non-dispersive Alfv\'en/ion-cyclotron
waves which are assumed to be injected at the left footpoints of the strands
and propagate along the field lines. A power-law spectrum is assumed for these
waves at the left footpoint, $s=0$, such that
\begin{equation}
\mathfrak{B}_{\omega}^{i}(s=0)=\alpha_{i}\left[\omega\ln(\Gamma^{2})
\right]^{-1}\mbox{erg cm$^{-3}$s$^{-1}$},\mbox{ \ \ \ \ i = 1, 2..7 }
\label{eq.1}
\end{equation}
where $\mathfrak{B}_{\omega}^{i}$ denotes the wave spectral energy density.
The parameter $\alpha_{i}$ which corresponds to the footpoint wave intensity
of each loop strand is listed in Tab.~\ref{6tbl-1}

The heating associated with the dissipation of the Alfv\'{e}n waves via
resonant wave-particle interactions is a fully kinetic process and does not
involve collisions (which are rare in solar corona) for the dissipation of
the waves. In such process, the ions have to obey the cyclotron resonance
condition, $\omega-k_{\parallel}v_{\parallel}-\Omega_{i}=0$, with the parallel
wave vector $k_{\parallel}$, wave frequency $\omega$, and ion gyrofrequency
$\Omega_{i}(s)=q_{i}B(s)/m_{i}$, which varies along the strand with the
spatial coordinate $s$ through the magnetic field strength $B(s)$.
Here $q_i$ is the charge of ion species $i$ and $m_i$ its mass.

Since the cross section of the magnetic flux tube (chosen for each strands)
expands by more than twice its footpoint value up to the middle of the loop
strand, the waves can be strongly damped at distances not far from the left
footpoint, and thus heat the protons at different positions near the left
footpoint (leading to quasi-footpoint heating). More details on this
dissipation mechanism are given in previous papers
(\citet{Vocks2002b}, BR2008a).

\begin{table}[!t]
\caption{The parameter $\alpha_{i}$ corresponds to the wave power density
of strand "$i$", defined by its maximum proton temperature shown in Fig.~1.}
\label{6tbl-1}
\centering{}\begin{tabular}{|c|c|c|}
\hline
Strand  & $T_{{\rm {Max}}}$ (MK)  & $\alpha_{i}\times10^{5}$ \tabularnewline
\hline
1  & 0.7  & 0.5\tabularnewline
2  & 0.84  & 0.7 \tabularnewline
3  & 1.2  & 1.1 \tabularnewline
4  & 1.4  & 1.4\tabularnewline
5  & 1.66  & 1.9 \tabularnewline
6  & 1.95  & 2.5 \tabularnewline
7  & 2.1  & 3.0 \tabularnewline
\hline
\end{tabular}
\end{table}
All the loop strands have the same boundary conditions, i.e., their electron
densities at the footpoints ($s=0,s=L$) are given by
$N_{e}(s=0)=N_{e}(s=L)=5\times10^{9}$~cm$^{-3}$, and their electron
temperatures by $T(s=0)=T(s=L)=2\times10^{5}$~K.

For the sake of simplification we considered the lower transition region (or the
upper chromosphere) to be the boundary for our modelled loop. This is
justified since at these altitudes the plasma can be assumed to be still fully
ionized, and therefore we avoid the complexity of considering the neutral
part of the solar atmosphere. Also, the heating mechanism acting on the base
of the corona might be different from the one proposed here for coronal loops.
Recently, it was for example suggested that dissipation of low-frequency
Alfv\'{e}n waves could help in overcoming the low initial temperature in the
lower transition region \citep{Bourouaine2008c}. Furthermore, the universal
upflows driven by spicules (or chromospheric jets) can play a significant
role in supplying the lower corona with hot plasma \citep{Depontieu2009}.

At $t=0$, we consider as initial condition that all strands are in hydrostatic
equilibrium with a similar initial state, i.e., the strands are cool with
isothermal temperature, $T=0.2$~MK and have relatively small apex densities.
After about ten thousand seconds all the strands reach their steady final
states, and they get filled with hot plasma due to the evaporation process
and plasma acceleration, and thus form an unresolved TRACE or SOHO/EIT loop.
It was shown in BR2008a that the wave energy flux density needed to heat a
loop strand to coronal temperature is about $10^{5}$~erg~cm$^{-2}$~s$^{-1}$.

\begin{figure}[t]
\includegraphics[width=14cm,height=13cm]{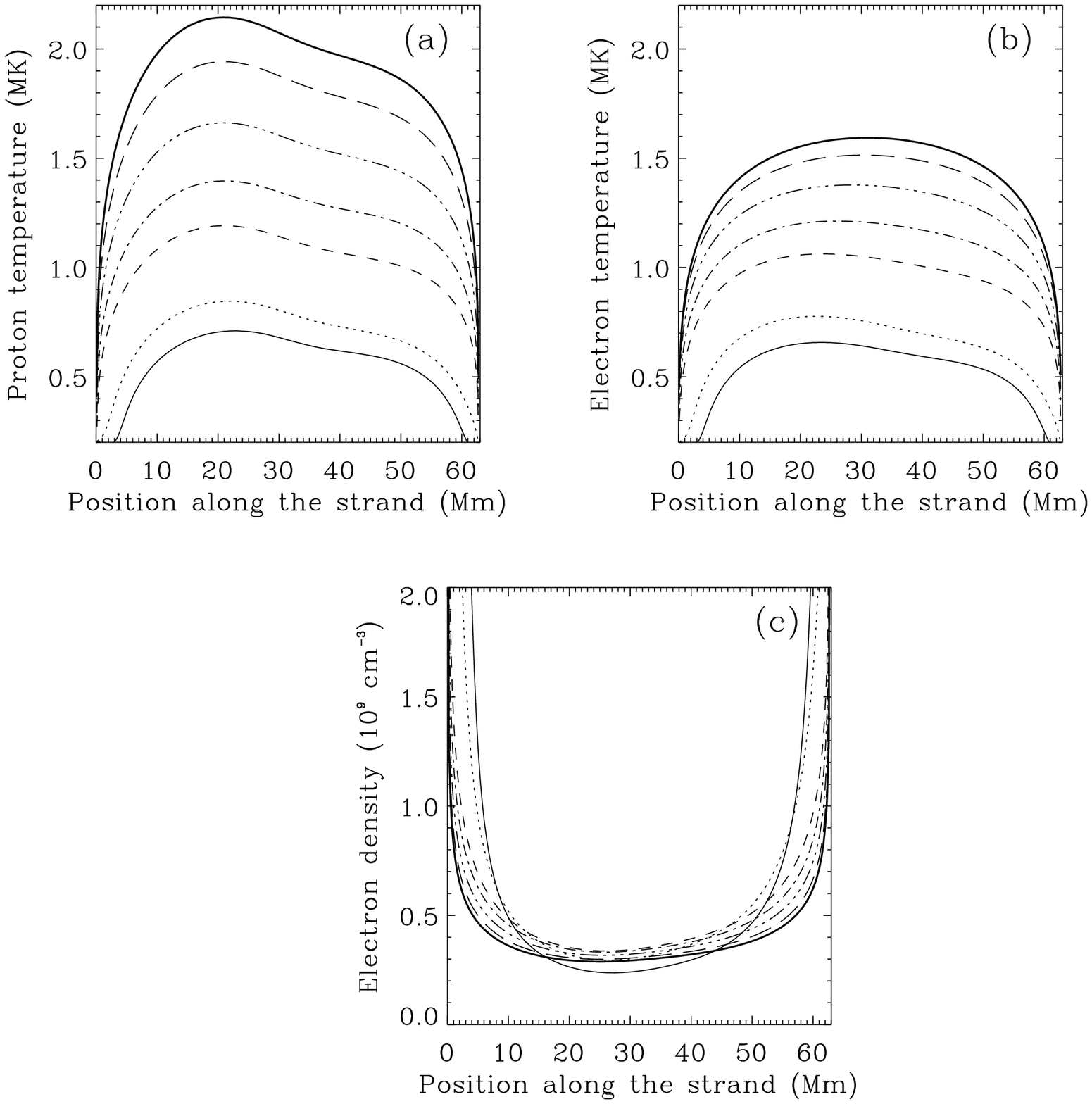}
\caption{Plasma profiles for each loop strand, (a): Proton temperature, (b)
electron temperature and (c) electron density. All parameters are
plotted as functions of strand position. The line style (solid line)
refers to strand "1", (dotted line)
strand "2", (dashed line) strand "3",
(dash-dotted line) strand "4", (dash-triple
dotted line) strand "5", (long dashed
line) strand "6" and (thick solid
line) strand "7". }
\label{fig.2}
\end{figure}

The temperature and density profiles of each loop strand are plotted in
Fig.~\ref{fig.2}. As expected, the higher the wave energy input the more
energy is transferred to the plasma loop strand. The Alfv\'en waves heat the
protons via ion-cyclotron-wave absorption, and thus due to proton-electron
collisions also the electrons can be heated. The proton temperatures increase
up to their maximum values close to a distance of $\approx20$~Mm where the
dissipation of the waves ceases. The proton temperatures tend to decrease
under the effect of heat conduction and electron-proton energy exchange by
collisions. However, the electron temperature is roughly equal to the proton
temperature as long as the latter is smaller than one mega kelvin. But when
the proton temperature exceeds this value, the electron temperature remains
at smaller values due to strong electron heat conduction. Therefore, the
electron and proton temperatures in the strands "1, 2" roughly overlap.
While, in the case of strands "3, 4, 5, 6, 7", the protons are hotter
than the electrons, and electrons show a quasi-uniform heating.

In this kinetic model, the loop strands can be filled with plasma via two
mechanisms, wave-induced particle acceleration and chromospheric plasma
evaporation (caused by heat conduction from the corona). However, wave-particle
acceleration is the dominant process to fill the loop-like structures with
protons, since the proton heat conduction is small, and the electron inertia
is neglected in our approximation. Thus, the plasma dynamics of the loop is
only related to the proton inertia. As a result, we see that the densities
in big parts of the loop strands do not differ by much. Consequently, the
strands have a roughly similar emission measure ($EM=N^{2}w$) across the loop.

Therefore, the modelled coronal loop (represented by a multi-strand loop)
has a varying cross-field electron temperature, which ranges from $0.6$ to
$1.6$~MK at its top, but only a small cross-field density variation (or
slight cross-field emission-measure variation). The next section is devoted
to the analysis of the emission of the modelled coronal loop, as it would be
observed in the three EUV channels at 171, 195 and 284~{\AA} available in
the TRACE imager or SOHO/EIT.

\section{Filter-ratio analysis}

In the following section we introduce the term ''isothermal temperature",
$T_{{\rm {iso}}}$, which simply means the observationally inferred loop
temperature, when assuming a single cross-field temperature and using the
filter-ratio technique.

\begin{figure}[!t]
\centering \includegraphics[width=14cm, height=6cm]{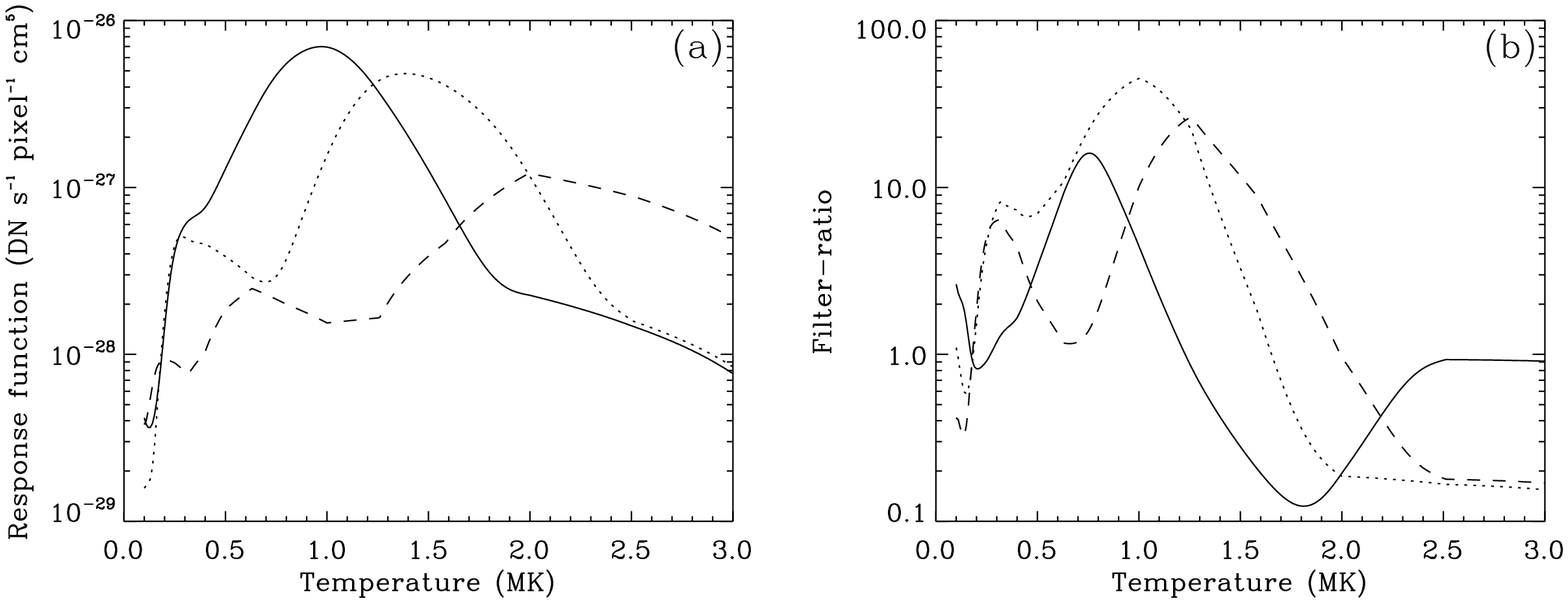}
\caption{(a): The response function for each TRACE filter; filter with 171~{\AA}
passband (solid line), filter with 195~{\AA} passband (dotted line) and filter
with 284 \AA{} passband (dashed line). (b): The ratio between two filters;
(171:195) (solid line), (171:284) (dotted line) and (195:284) (dashed line).}
\label{fig.3}
\end{figure}

It is possible to synthesize the total emission of the modelled coronal
loop from the summed strand emissions. Each strand intensity can easily
be computed from its density and temperature profile by using the following
emission measure relation:
\begin{equation}
I_{i}(s)=w_{i}(s)N_{i}^{2}(s)G^{\lambda}\left(T_{i}(s)\right),\mbox{ \ \ \ \ i = 1, 2..7 }
\label{eq.2}\end{equation}
where $I_{i}(s)$ is the intensity of the strand "i" given per unit optical
length (units of DN s$^{-1}$ pixel$^{-1}$) at position $s$ along the strand.
The symbols $T_{i}$ and $N_{i}$ denote, respectively, the strand electron
temperature (in units of kelvin) and density (in units cm$^{-3}$), and
$G^{\lambda}(T)$ is the response function for each filter centered around
wavelength $\lambda$; it is plotted in frame (a) and the filter ratio in
panel (b) of Fig.~\ref{fig.3}.

\begin{figure}
\includegraphics[width=14cm,height=13cm]{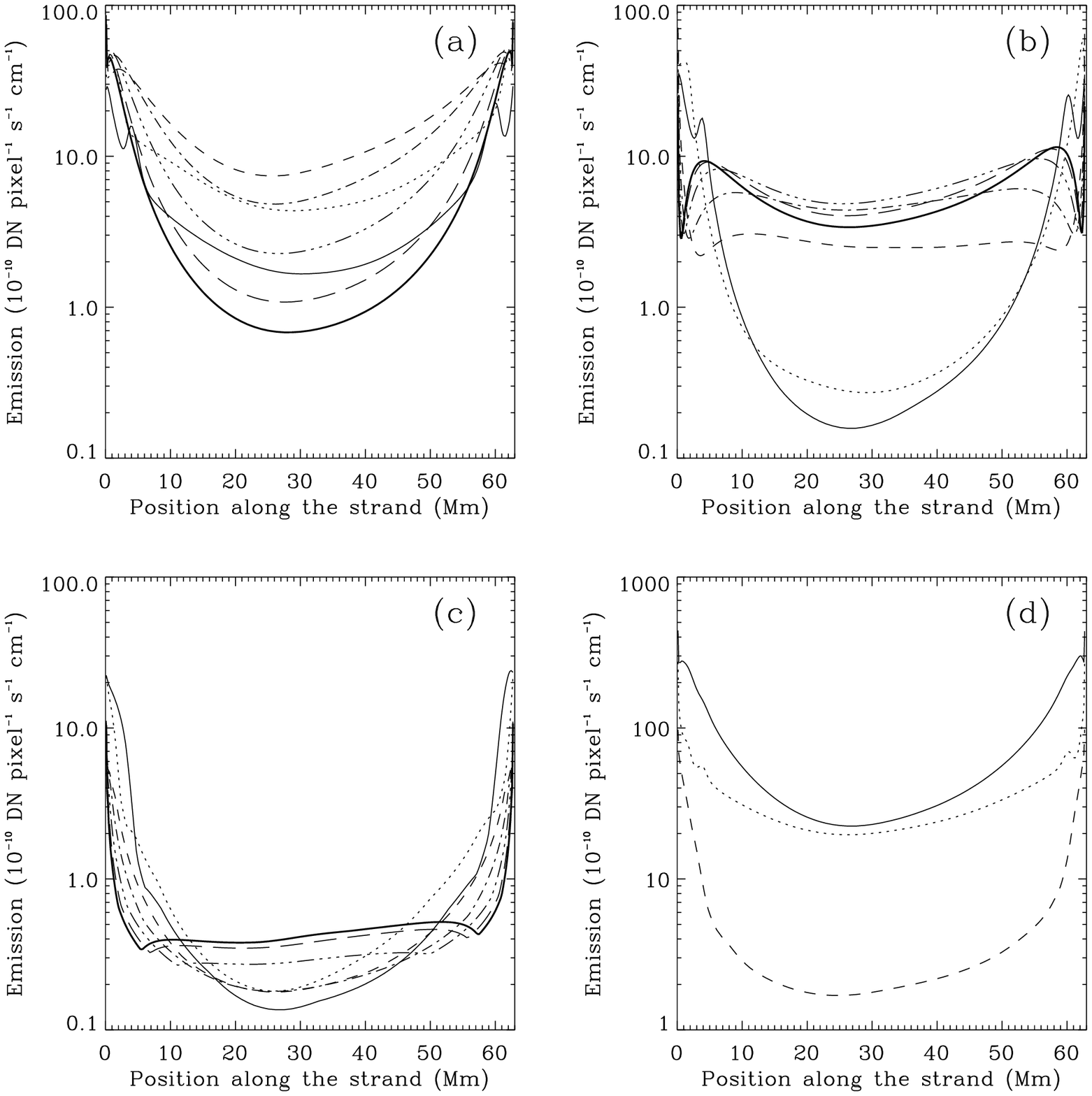}
\caption{(a): Emission of each strand through the filter (171~{\AA}). (b):
Emission of each strand through the filter (195~{\AA}). (c): Emission
of each strand through the filter (284~{\AA}). All parameters are plotted as
a function of the strand position. Each strand is represented by lines
of the same style as used in Fig.~\ref{fig.2}. (d): Total loop emission
through the filters (171~{\AA}) (solid line), (195~{\AA})
(dot line) and (284~{\AA}) dashed line.}
\label{fig.4}
\end{figure}

The strand emissions, $I_{i}/w_{i}$, for the three filters are plotted
in Fig.~\ref{fig.4}. It seems that the brightest strands in the 171~{\AA}
emission are those labelled with numbers 3, 4, and 2 (ordered from the highest
to lowest emission). The 171~{\AA} filter is more sensitive to the strands
having temperatures from 0.7~MK to 1.2~MK. Therefore, the first three strands
mainly contribute to the total emission of the coronal loop in the 171~{\AA}
passband (see Fig.~\ref{fig.4}d). Moreover, from Fig.~\ref{fig.4}b it seems
that the 195~{\AA} filter is sensitive to all strands, except for
those labelled with numbers 1 and 2. This is because this filter is sensitive
to the plasma temperatures from 1~MK to 1.4~MK. However, less visibility of
the loop strands is expected in the filter 284~{\AA} (see Fig.~\ref{fig.4}c),
since this filter is sensitive only to hot plasma with $T\geq1.6$~MK,
and thus, the total loop also appear with a small emission in this filter
(see Fig.~\ref{fig.4}d). Furthermore, in the realistic case when the
background has to be accounted for, this modelled loop can hardly be seen in
the 284~{\AA} passband, since its emission would be contaminated strongly by
background effects. Therefore, we can classify our modelled multi-strand loop
as a warm EUV loop ($T\leq1.5$~MK) in terms of visibility in the TRACE
or SOHO/EIT filters.

\begin{figure}
\includegraphics[width=14cm,height=13cm]{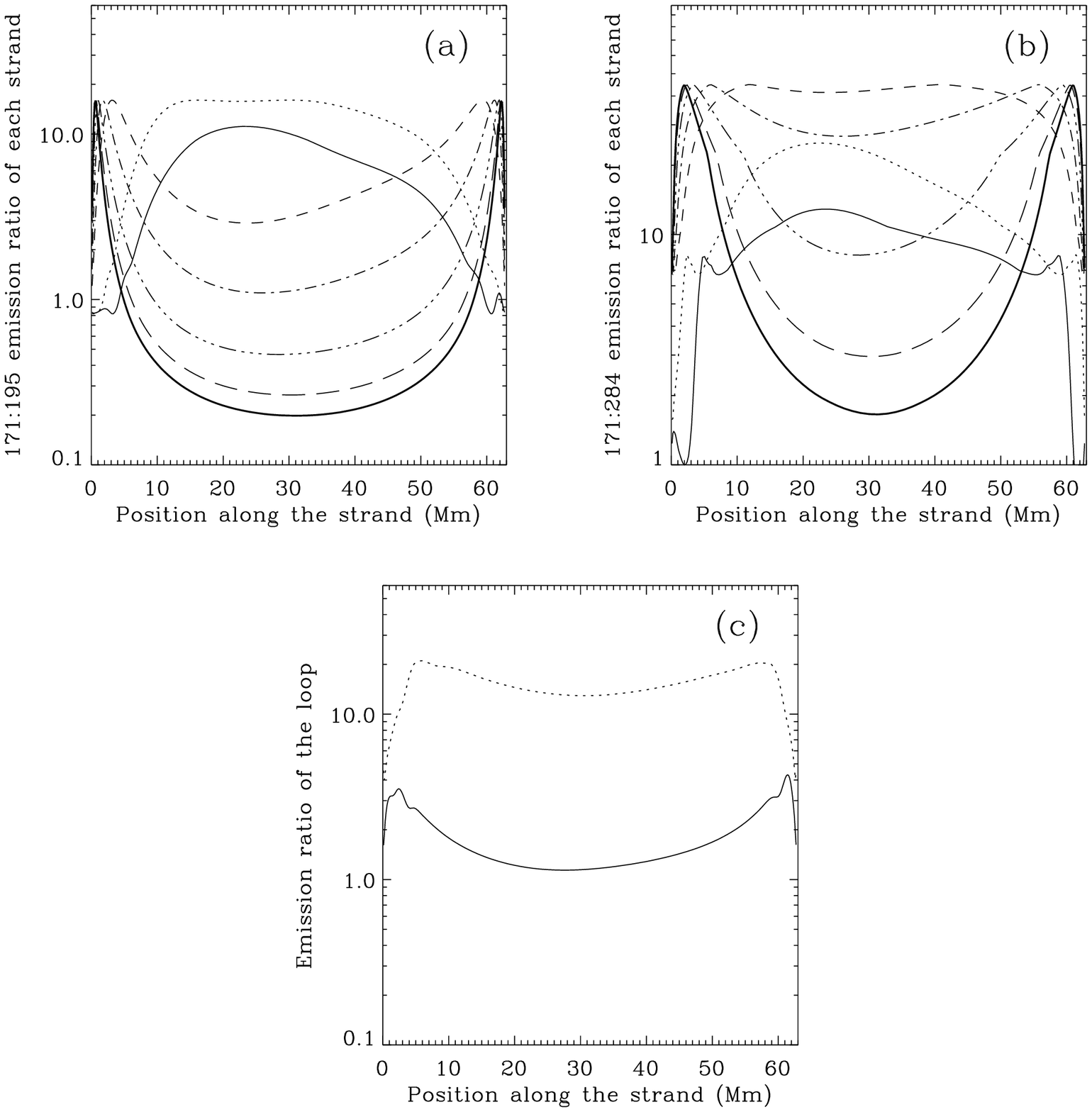}
\caption{Emission ratios 171:195 (a), 171:284 (b), plotted for each strand
represented by the same type of line as given in Fig.~\ref{fig.2}.
(c): Averaged emission ratios of the loop, (171:195) (solid line)
and (171:284) (dot line).}
\label{fig.5}
\end{figure}

Fig.~\ref{fig.5}c displays the emission ratios of the modelled coronal
loop versus its loop length. Interestingly, the emission ratio 171:195
(indicated by a solid line) varies slightly between two values, 1
and 3 inside the loop. However, the emission ratio 171:284 (indicated
by dot lines) spans between the values between 10 and 12. As we will
show, the variations in these two emission ratios are still not large
enough to produce variations in the isothermal temperatures along
the loop length.

\begin{figure}
\includegraphics[width=14cm,height=6cm]{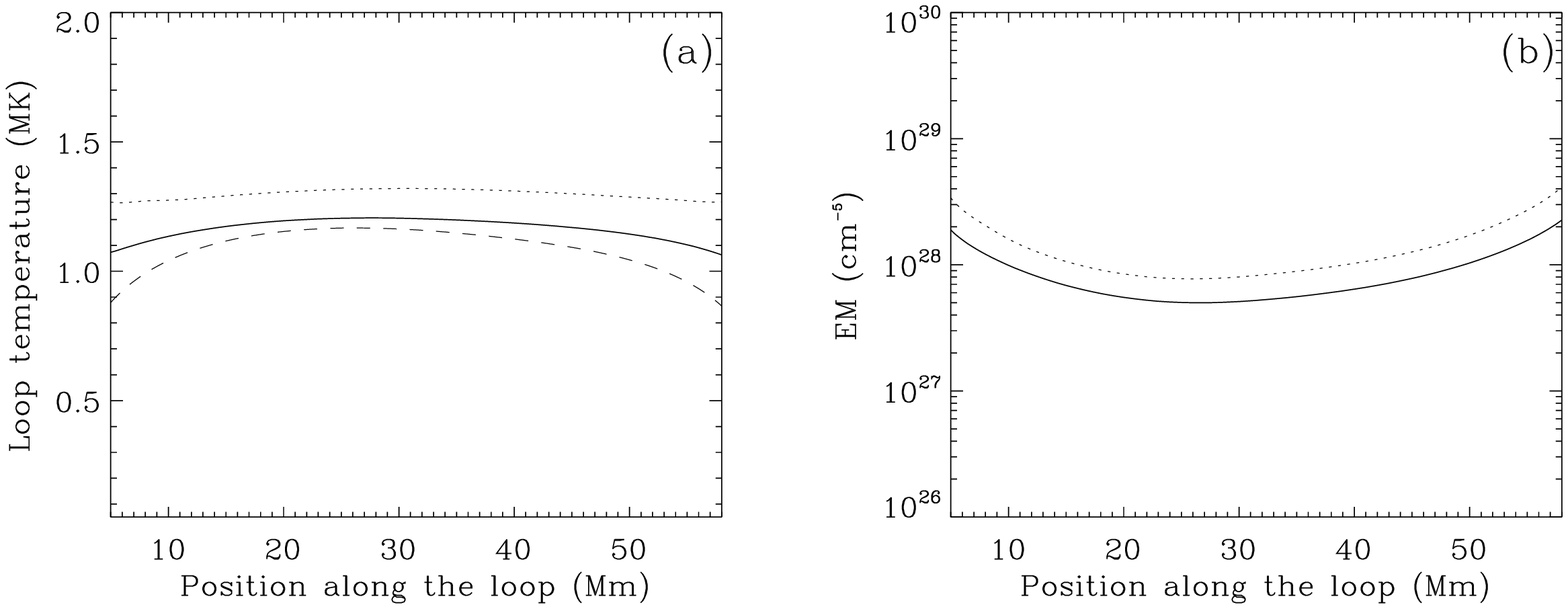}
\caption{(a) and (b): Loop temperatures and loop emissions. (a): Isothermal
temperature, $T_{iso1}$ that corresponds to the filter ratio (171:195)
(solid line), $T_{iso2}$ that corresponds to the filter ratio (171:284)
(dotted line) and the average temperature (dashed line). (b): Emission
measures correspond to filter ratios (171:195) (solid line) and (171:284)
(dotted line). The line of sight depth assumed here is $10^{10}$~cm,
a value suggested by \citet{Lenz1999} to compute the emission measure.}
\label{fig.6}
\end{figure}

Based on the obtained emission ratios, and by making use of the ratios
between the response functions plotted in Fig.~\ref{fig.3}b, we
can extract the isothermal temperatures that would be inferred from
TRACE or EIT observations of our modelled coronal loop. The inferred
loop temperature profiles from the two filter ratios, 171:195 and
171:284, are plotted in Fig~\ref{fig.6}a. Notice that these inferred
temperatures differ from the computed averaged temperature profile.

It turns out that there are two distinct temperature profiles that can be
inferred from the filter-ratio analysis for a simulated TRACE or EIT coronal
loop. This is a consequence of the multi-thermal property of our modelled
coronal loop. It seems as if the use of two different filter ratios in the
TRACE or EIT coronal loop observations allows us to determine whether these
observed loops are isothermal or multi-thermal. If the two filter ratios do
not provide the same temperature profile, this may be an indication that the
cross-field single temperature assumption was not satisfied. This conclusion
may explain the results obtained by \citet{Schmelz2003}. The authors analysed
10 coronal loops that were clearly visible in the 171, 195 and 284~{\AA}
passbands of the EIT. They showed that in each case of the used background
substraction method two different uniform temperatures were obtained, one
from the 171:195 ratio and the second for the 195:284 ratio. The authors
suggested that the single cross-field temperature assumption could provide a
misleading loop temperature.

However, we argue here that, in case of warm loops ($T\leq1.5$~MK), it is
difficult to infer the temperature from the filter ratio 195:284. This is
because the ratio function 195:284 in Fig.~\ref{fig.3}b peaks at a
temperature $T\sim1.26$~MK, and for any ratio value that spans between
$\sim1$ and $\sim26$, there are two corresponding temperatures; one is above
1.26~MK and the other below it (non-unique solution). Therefore, in the case
of warm coronal loops, it is difficult to constrain the temperature value to
be chosen. For that reason, we did not use here the filter ratio 195:284 to
determine the loop temperature.

By assuming a line-of-sight depth of $10^{10}$ cm, as considered
by \citet{Lenz1999}, we estimate the emission measure from the filter
ratio (171:195) and (171:284) (see Fig~\ref{fig.6}b). The emission
measure profile implies an enhanced density inside the loop, and the
estimated EM for the first filter ratio is comparable to the one obtained
by \citet{Lenz1999}, however, a higher EM could be obtained if we
considered the filter ratio (171:284).

For warm coronal loops, the use of the triple filters 171/195/284~{\AA}
may not help to decide if these loops are isothermal across the field
or not. This is because these loops may not appear clearly in the filter
(284~{\AA}) \citep{Chae2002}. Their emission can be strongly affected by the
background emission, and therefore, if their appearance in the filter
(284~{\AA}) leads to a temperature uncertainty of the order of ten percent,
then it would be difficult to discriminate between the two temperature
profiles obtained from the two filter ratios (171:195) and (171:284).

\begin{figure}
\includegraphics[width=14cm,height=13cm]{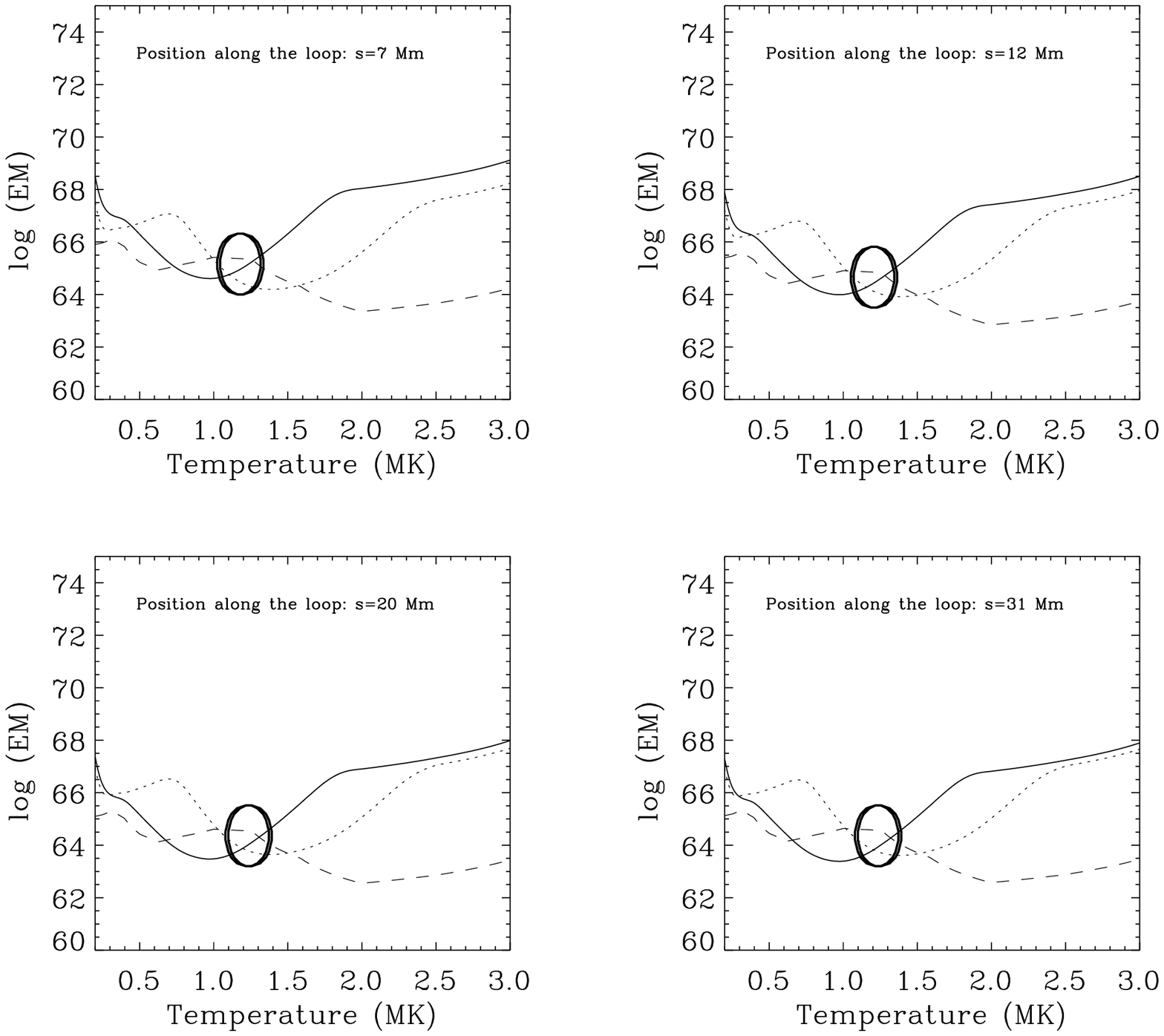}
\caption{Loop emission measure curves for (171~{\AA}) (solid line), filter
(195 \AA{}) (dotted line) and filter (284~{\AA}) (dashed line). The
line of sight depth assumed here is $10^{10}$~cm, as was suggested
by \citet{Lenz1999}.}
\label{fig.7}
\end{figure}

Fig.~\ref{fig.7} displays EM-loci plots at some position along the modelled
loop. It is clear that the curves do not exactly intersect in a single point.
However, these curves are approaching each other in a narrow temperature
interval, and in some papers \citep{Schmelz2009,Tripathi2009} the authors
suggested that this narrow crossing area in the EM curves is due the not
strictly isothermal nature of the loop along the line of sight. In the
general case, this also can be a misleading interpretation. Although the
modelled loop is completely multi-thermal, the curves only intersect in a
narrow temperature interval around $T\sim1.2$~MK.

For example, by considering the photometric uncertainties determined by
\citet{Patsourakos2007}, our simulated coronal loop is found to be consistent
with an isothermal plasma. The authors define a new quantity based on
triple-filter analysis called, the average logarithmic difference, which is

\begin{equation}
\Delta\log[EM(T)]=\frac{1}{3}\{\log\frac{EM_{171}}{EM_{195}}
+\log\frac{EM_{171}}{EM_{284}}+\log\frac{EM_{195}}{EM_{284}}\}.
\end{equation}

This quantity considers all three of the possible channel combinations. It
vanishes when the emission measures, obtained from the three possible
filter-ratios, are identical at the isothermal temperature, $T_{iso}$.
This is the case for accurate observations of an isothermal plasma.
However, in real measurements the errors have to be account for, which
are characterized by an uncertainty $\Delta\log(EM)_{uncertain}$.

The uncertainty in each emission-ratio is obtained from the determination of
the uncertainty in the corresponding intensity-ratio. The intensity
measurements are affected by systematic errors due to imperfect knowledge of
the photometric calibration, and therefore, the uncertainties associated with
relative photometry errors among the 171, 195, and 284~{\AA} channels have to
be characterized. \citet{Patsourakos2007} have estimated the average
uncertainty for all filter combinations, $\Delta\log(EM)_{uncertain}$, to be
$0.06$. However, after a CDS/TRACE/EIT inter-calibration study in a quiet-Sun
region, \citet{Brooks2006} found that the 284~{\AA} filter sensitivity could
be underestimated by a factor of approximately 3, and therefore, the
uncertainty $\Delta\log(EM)_{uncertain}$ could reach the value 0.63.

\begin{figure}[!t]
\centering \includegraphics[width=14cm, height=13cm]{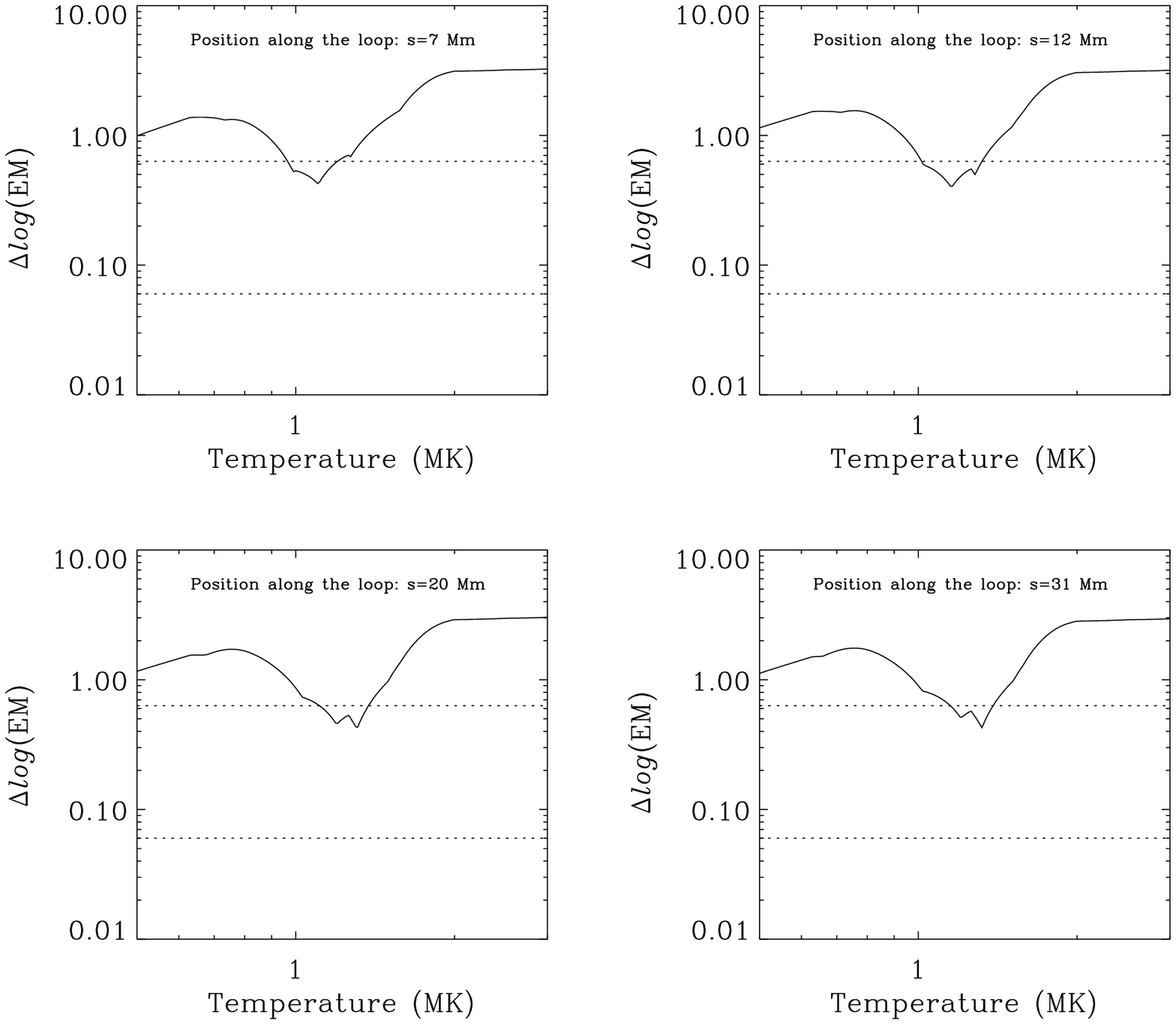}
\caption{Average logarithmic difference of the isothermal emission measures versus temperature.
The horizontal dashed lines correspond to the lower, respectively the upper, limit of the uncertainty
of the average logarithmic difference as determined by \citet{Patsourakos2007}.}
\label{fig.8}
\end{figure}

The isothermal interpretation may only be valid when $\Delta\log(EM)$
is smaller than $\Delta\log(EM)_{uncertain}$. In Fig.~\ref{fig.8}
we plot the $\Delta\log[EM(T)]$ curve at some locations along the
loop. There are absolute minima near $1.2-1.4$~MK which are below
the error bar. As we can see, although our simulated loop is multi-thermal,
the loop can be considered as isothermal according to the argument
given above. This means that with the triple-filter method suggested
by \citet{Patsourakos2007}, it is not possible to confirm the isothermal
interpretation assumed for the observed loops.

There is an interesting behaviour in the inferred temperature profiles
of our modelled coronal loop, which is that these profiles are quasi-uniform
along the loop length (see Fig.~\ref{fig.6}a). This notion seems to fit the
TRACE coronal loop features as discussed in the introduction.

According to our modelled loop,  the flatness in the temperature profile
is consequence of loop consisting of filaments, which have similar emission
measures, $EM_{i}$ and different temperatures in the cross-field direction.
In such case, the overall loop intensity ratios, at a given loop position,
simply would reduce to the ratios of the response functions summed over the
number of the loop strands, $N_{s}$, within this coronal loop position, i.e.,
to the ratios
\begin{equation}
R  =  \frac{\sum_{i=1}^{N_{s}}G^{\lambda_{1}}(T_{i})}{\sum_{i=1}^{N_{s}}G^{\lambda_{2}}(T_{i})},
\end{equation}
where, $\lambda_{1}$ and $\lambda_{2}$ are the wavelengths of considered filters.

Therefore, flat temperature profiles, which are often obtained when using
the filter-ratio technique, can even occur in the case when the loop consists
of a small number of strands (forming a real discrete loop structure) with
features described above. However, \citet{Weber2005} suggested that the
flatness of the temperature profile may also occur if a loop has continuum
cross-field structure of varying cross-field temperature (i.e., constant
broad differential emission measure (DEM)). In such a case, the emission ratio
method is biased toward the ratio of the integrated response functions over
a broad temperature interval. This is nothing but a particular example of
equation (4), when assuming a large number of strands (forming a continuum
structure).

\section{Conclusion}

In this work we synthesized the emission of seven strands which together
are assumed to constitute a coronal loop as seen in a low-resolution
TRACE or SOHO/EIT observation. Each of the loop strands used in the model
is heated by Alfv\'en/ion-cyclotron waves via wave-particle interactions.
This process leads to proton heating in the electron-proton plasma,
and due to electron-proton collisions the electrons can then be heated
as well up to a certain temperature which is below or equal to the
proton temperature. It turns out that the plasma, in case of hot loop
strands (where the proton temperature exceeds $>1$~MK) is not in
local thermal equilibrium and the electrons are cooler than the protons.

In our model, the Alfv\'en waves, which are assumed to penetrate the strands
from their footpoints, are there generated with different intensities.
Consequently, different heating profiles occur within each strand due to the
wave absorption or heating process. Therefore, this differential heating
leads to a varying cross-field temperature in the total coronal loop. Moreover,
due to the ion acceleration process caused by wave-particle interactions, the
model produces roughly similar electron density values (or emission measure
since the strands have the same widths) across the loop.

The simulated TRACE observation of the modelled loop implies two different
quasi-uniform temperature profiles along the loop length, one derived from
the filter-ratio 171:195 and the other for the 171:284. This flatness behaviour
in the temperature profiles is consistent with results based on the filter
ratio analysis applied to TRACE/EIT coronal loop. According to our model
results, such flat temperature profiles can occur when the loop consists of
a small number of threads (forming a discrete loop like-structure) which
have different temperature profiles but rough similar emission measure, $EM$,
across the total loop.

It is worth noting that, due to the week binary collisions between the plasma
species, the loop plasma can be far from local thermal equilibrium if the ion
temperature exceeds one mega Kelvin. Therefore, it is ideally suited for
applying the kinetic description of a plasma rather than using the single
fluid approach to model the coronal loop heating.

Furthermore, most of the previous hydrodynamic models invoked arbitrary
heating functions in the energy equation to model the loop heating. However,
our model adopts a physical mechanism for loop heating. The considered heating
process relies on wave-particle interactions described by the quasi-linear
theory of the kinetic Vlasov equation. This process does not only lead to
plasma heating (or ion heating) but also accelerates the ions in the loop
strands due to the wave pressure, and thus, can fill these strands with hot
plasma. This loop-filling mechanism, which is different from the evaporation
process, is naturally involved in our model, and it turns to be important
in order to reproduce the observed features of coronal EUV loops.

\end{document}